# Production of $D_s^\pm$ mesons in Au+Au collisions at $\sqrt{s_{NN}}$ = 200 GeV by STAR

**Chuan Fu (for the STAR collaboration)**[*]

*Central China Normal University,*
*No.152 Luoyu Road, Wuhan, China*

*E-mail:* fuchuan@mails.ccnu.edu.cn

Charm quarks are an excellent probe to study properties of the Quark-Gluon Plasma (QGP) created in ultra-relativistic heavy-ion collisions. In particular, measurements of the $D_s^\pm$ meson production can provide valuable information on the charm quark hadronization mechanism in the QGP. We report the measurements from the STAR experiment on the invariant yields of $D_s^\pm$ mesons as a function of transverse momentum for different centrality classes of Au+Au collisions at $\sqrt{s_{NN}}$ = 200 GeV. These measurements utilize the data with the Heavy Flavor Tracker detector from 2014 and 2016. The ratio between strange ($D_s^\pm$) and non-strange ($D^0$) open charm mesons will also be shown, and compared to PYTHIA and different model calculations. A clear enhancement relative to the PYTHIA calculation is seen in the ratio, while different model calculations incorporating strangeness enhancement and charm quark coalescence hadronization can describe the observed enhancement qualitatively. These results suggest that recombination of charm quarks with strange quarks in the QGP plays an important role in charm quark hadronization.

---

[*]Speaker

**1. Introduction**

A new state of matter, Quark-Gluon Plasma (QGP), is formed under extremely high temperature and energy density at Relativistic Heavy Ion Collider (RHIC), in which quarks and gluons are released from their confinement inside of hadrons. Heavy quarks are predominately created from initial hard scatterings since their masses are larger than the temperature of QGP, and experience the entire QGP evolution. Therefore, heavy quarks are considered an excellent probe to study the properties of QGP. In the QGP medium, $D_s^+$ ($D_s^-$) mesons which consist of a charm (anti-charm) quark and an anti-strange (strange) quark are argued to form by recombination of charm quarks and strange quarks. $D_s^\pm$ meson yield is expected to be increased due to enhanced strangeness density in Au+Au collisions relative to p+p collisions. Theoretical calculations predict that the $D_s^\pm/D^0$ yield ratio in Au+Au collisions will show an enhancement compared to that in p+p collisions [1].

**2. Data set and analysis**

About two billion minimum bias (MB) triggered events collected from year 2014 and 2016 are used in this analysis. The Heavy Flavor Tracker (HFT) was operating during the data collection. The HFT consists of three sub-systems: innermost two layers of Pixel detectors, the intermediate Silicon Tracker and outermost layer of Silicon Strip Detector. The excellent pointing resolution (< 30 $\mu m$ when track momentum > 1.0 GeV/c) provided by the HFT allows to reconstruct decay vertices of charm hadrons, which greatly improves the signal-to-background ratio for charm hadron reconstruction. The trajectories and momenta of charged particles are reconstructed via the Time Projection Chamber (TPC) and HFT. The $D_s^\pm$ are reconstructed by the decay channel of $D_s^\pm \to \phi\, (\phi \to K^+ + K^-) + \pi^\pm$. The flight time recorded by the Time Of Flight (TOF) detector and measured ionization energy loss in the TPC provide particle identification. To improve the signal significance, the Boosted Decision Tree (BDT) from Toolkit for MultiVariate Analysis (TMVA) [2] was applied. The BDT response is evaluated by the BDT classifer which is constructed from variables characterizing decay topology. The $D_s^\pm$ signal with the best significance is obtained by optimizing the BDT cut for each centrality and transverse momentum ($p_T$) bin. The reconstruction efficiency is calculated from data-driven simulation [3] by applying the same acceptance cuts, TPC tracking cuts, TPC-to-HFT matching, particle identification cuts, and topological variable cuts. The $D_s^\pm$ invariant yield, $((1/2\pi p_T)d^2N/dp_T dy)$ is obtained as the average yield per event scaled by the inverse of reconstruction efficiencies and the decay branching ratio.

**3. Results**

The invariant mass distribution of $M_{K^+K^-\pi^\pm}$ from 0-80% collision centrality is shown in Fig.1 (left panel). The black solid line depicts a two-Gaussian fit for $D_s^\pm$ and $D^\pm$ signal plus a linear fit for background. The raw signal yields are obtained by subtracting the combinational backgrounds, evaluated by integrating linear function (blue dotted line), from the number of candidates within 3 standard deviations from the $D_s$ mass. Figure 1 (right panel) shows the $D_s^\pm$ invariant yields as a function of $p_T$ for 0-10%, 10-40% and 40-80% collision centralities. Vertical bars (smaller then markers and not visible) and brackets on data points depict the statistical and systematic uncertainties, respectively. The measurement reaches to low $p_T$ (~1.0 GeV/c), which is crucial for extracting the total $D_s^\pm$ production cross section.

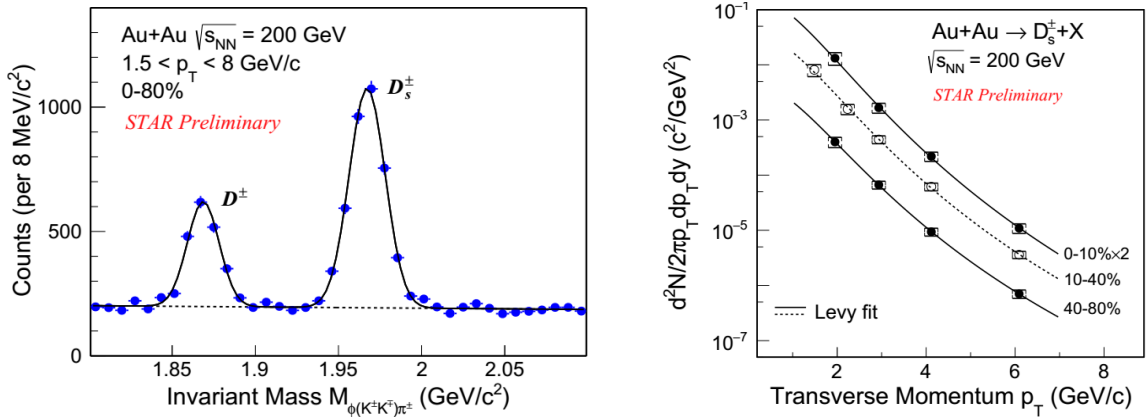

**Figure 1:** (left panel) The invariant mass distribution $M_{K^+K^-\pi^\pm}$ in 0-80% Au+Au collisions at $\sqrt{s_{NN}}$ = 200 GeV. (righ panel) The $D_s^\pm$ invariant yield as a function of $p_T$ in various centrality bins in Au+Au collisions at $\sqrt{s_{NN}}$ = 200 GeV. Solid and dashed lines represent Levy function fits in each centrality bin.

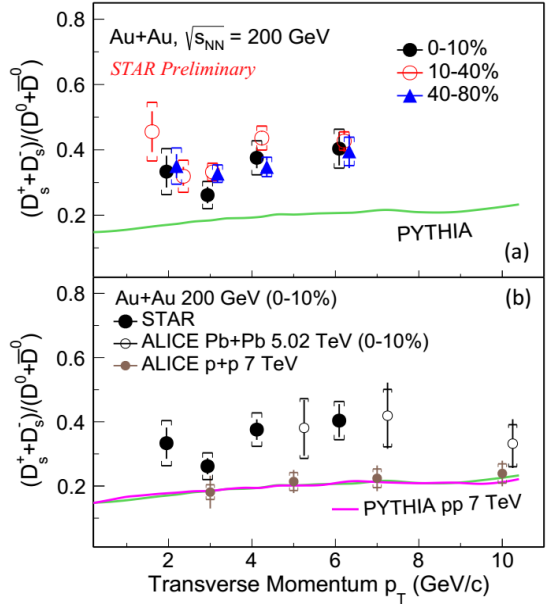

**Figure 2:** (a) The $D_s/D^0$ yield ratio as a function of $p_T$ in different centrality bins of Au+Au collisions at $\sqrt{s_{NN}}$ = 200 GeV compared to a PYTHIA calculation for p+p collisions at same energy. (b) The STAR measurement of $D_s/D^0$ yield ratio as a function of $p_T$ in 0-10% central Au+Au collisions at $\sqrt{s_{NN}}$ = 200 GeV compared to the ALICE measurements in Pb+Pb and p+p collisions and PYTHIA calculations.

Figure 2 (a) shows the $D_s/D^0$ yield ratio as a function of $p_T$ for different centrality bins in Au+Au collisions by STAR and the PYTHIA calculations [4] for p+p collisions. The STAR measurements show a large enhancement (about a factor of 1.5-2.0) compared to PYTHIA calculation and exhibit a weak $p_T$ dependence. Figure 2 (b) shows the $D_s/D^0$ yield ratio for 0-10% centrality bin from the STAR measurement and Pb+Pb collisions at $\sqrt{s_{NN}}$ = 5.02 TeV from ALICE in the same centrality bin [5]. They are compatible within uncertainties in the overlapping $p_T$ region. The $D_s/D^0$ yield ratios from ALICE p+p collisions [6] and PYTHIA calculation at 7 TeV are also shown for comparison. The ALICE p+p results are systematically lower than those in heavy-ion collisions and can be well described by the PYTHIA calculation.

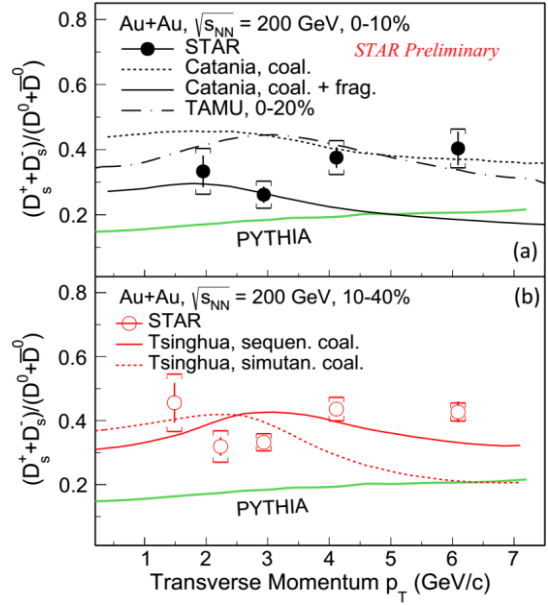

**Figure 3:** The $D_s/D^0$ yield ratio as a function of $p_T$ compared to model calculations in 0-10% (a) and 10-40% (b) collision centralities at $\sqrt{s_{NN}}$ = 200 GeV.

Several model calculations incorporating coalescence hadronization of charm quarks and strangeness enhancement are shown in Fig. 3 (a) and (b). The Tsinghua model [7] is the only model that considers sequential coalescence ($D_s^\pm$ mesons hadronize earlier than $D^0$), denoted by "Tsinghua, sequen. coal.". It does not include fragmentation hadronization. The Catania model [8] provides a calculation for coalescence only, "Catania, coal.", and a calculation including both fragmentation and coalescence, "Catania, coal.+frag.". The TAMU model [9] employs a resonance recombination and includes fragmentation in addition to coalescence. For the 0-10% centrality bin, the calculations from TAMU and Catania (coal.) describe the enhancement seen in data relative to PYTHIA calculation above 3.5 GeV/c. The "Catania, coal.+frag." calculation is consistent with data for $p_T <$ 3.5 GeV/c, but underestimates the measurement above 3.5 GeV/c. For the 10-40% centrality bin, it shows that the Tsinghua (sequen. coal.) model can qualitatively describe the data within uncertainties. These comparisons indicate that the coalescence hadronization plays an important role in charm quark hadronization in the QGP medium.

## 4. Summary

In summary, STAR reports the first measurement of $D_s^\pm$ production and $D_s/D^0$ yield ratio as a function of $p_T$ for different collision centralities at midrapidity (|y| < 1) in Au+Au collisions at $\sqrt{s_{NN}}$ = 200 GeV. The $D_s/D^0$ ratio shows a clear enhancement compared to that from PYTHIA

prediction. The enhancement can be described by model calculations incorporating charm quark recombination with enhanced strange quark in the QGP. These results indicate that charm quark coalescence plays an important role in hadronization.